\documentclass[epj]{svjour}
\usepackage{latexsym}
\usepackage{graphics}
\begin{document}
\title{Fractal geometry of Ising magnetic patterns: signatures of criticality and diffusive dynamics}
\author{E. Agliari\inst{1} \and R. Burioni\inst{1,2} D. Cassi\inst{1,2} \and A. Vezzani\inst{1,2}
}                     % Do not remove
\institute{Universit\`a degli Studi di Parma, Parco Area delle
Scienze 7/a, 43100 Parma, Italy \and Istituto Nazionale Fisica
della Materia (INFM), UdR PARMA, Parco Area delle Scienze 7/a,
43100 Parma, Italy}
\date{Received: date / Revised version: date}

\abstract {We investigate the geometric properties displayed by
the magnetic patterns developing on a two-dimensional Ising
system, when a diffusive thermal dynamics is adopted. Such a
dynamics is generated by a random walker which diffuses throughout
the sites of the lattice, updating the relevant spins. Since the
walker is biased towards borders between clusters, the
border-sites are more likely to be updated with respect to a
non-diffusive dynamics and therefore, we expect the spin
configurations to be affected. In particular, by means of the
box-counting technique, we measure the fractal dimension of
magnetic patterns emerging on the lattice, as the temperature is
varied. Interestingly, our results provide a geometric signature
of the phase transition and they also highlight some non-trivial,
quantitative differences between the behaviors pertaining to the
diffusive and non-diffusive dynamics.
  % end of PACS codes
} %end of abstract
\maketitle
\section{Introduction}
\label{intro} In an earlier paper \cite{earlier} we introduced a
thermal dynamics for an Ising ferromagnet exhibiting a diffusive
character: time evolution is performed by a walker which,
according to a local probability, can hop across the sites of an
underlying lattice and can possibly flip the relevant spins. In
order to make such a dynamics consistent with some physical
processes, which make the system evolve \cite{earlier,buonsante},
the walker is biased towards high energy regions.

Due to a very general structure, our algorithm works as well for
arbitrary lattices, made up of spins which can assume an
arbitrary, finite number of states.

%In fact, while the usual non-diffusive dynamics typically update
%the lattice by sweeping along its parallel lines, here, the walker
%is intended as a localized excitation diffusing throughout the
%whole sample. Moreover, since we expect such a localized
%excitation to occur more likely where a spin-flip is energetically
%more favorable, the walker diffusion is properly biased towards
%those sites. This kind of bias distinguishes our random walker
%from the number of different models which, in the last decades,
%have been introduced to be employed in several fields (physical,
%chemical, biological and even social sciences)
%\cite{montroll,weiss,haus,ordermann,masuda}.

The thermodynamic effects of this thermal dynamics have already
been explored \cite{earlier}, showing that, though the steady
states reached by the system are non-canonical, the universality
class is preserved. It was also noticed that the geometry of
magnetic patterns developing on the lattice could be affected by
the dynamics adopted. Indeed, due to the bias the walker is
endowed with, we expect sites pertaining to boundaries between
clusters to be more frequently updated. As a result, we also
expect to have differently-shaped spin arrangements, with respect
to a non-diffusive dynamics.

Hence, in this work, by inspecting the geometric properties of
magnetic patterns, we aim to further deepen the study of the
dynamics introduced. We underline that, while previous works
dealing with the geometry of magnetic configurations
\cite{janke,coniglio,antoniou,vanderzande,vanderzande2,deng,duplantier}
mainly concerned the analysis of the critical clusters (namely
clusters emerging when $T=T_c$), here we are interested in how the
diffusive character of the dynamics influences the evolution of
the spin arrangement on the whole lattice, as the temperature is
varied.

An appropriate parameter to quantitatively describe clusters, in
terms of geometric properties, is the fractal dimension
\cite{guenoun,han,monceau}. Among the several methods introduced
at this purpose \cite{janke,asikainen,block,dubuc,halsey}, we
select the box-counting technique which, as will be explained in
Sec.~\ref{geom}, let us to easily achieve a complete description
of the entire lattice. Moreover, as we will show, our geometric
measures provide a significant signature of the critical point
and, by a proper fit of data, we are also able to extract good
estimates for $T_c$.

The same measures are performed on both spin-1/2 and spin-1 Ising
systems, in order to achieve a wider insight into the matter and,
possibly, to extend results to the general spin-S case. As
discussed in the following, our results are strongly concerned
with the symmetry existing among Ising spin-states, and, in fact,
distributions of negative (positive) and null spins, display very
different behaviors. In particular, as approaching $T_c$,the
pertaining fractal dimension are described by a power and a linear
law, respectively. Qualitatively analogous results are also
obtained when a non-diffusive dynamics, with sequential
spin-updating, is adopted. However, while the diffusive dynamics
introduced recovers the canonical thermodynamics critical
exponents \cite{earlier}, as far the geometric exponents, some
interesting quantitative differences emerge.

The layout of the paper is as follows. In Sec.~\ref{dyn} we
explain how our diffusive dynamics works, while in Sec.~\ref{geom}
we deal with the geometrical features, especially focusing on the
critical range. Finally, Sec.~\ref{concl} contains a summary and a
discussion of results.

\section{Diffusive Dynamics}
\label{dyn} In this Section we resume how the diffusive dynamics
introduced works, while a detailed description can be found in
\cite{earlier}.

%It should be underlined that our dynamics is not meant to recover
%the canonical distribution, but rather to perform some physical
%processes which make the system evolve \cite{earlier,buonsante}.

Albeit our dynamics provides a versatile tool to be applied to an
arbitrary spin-S discrete system, in this work we are focusing on
spin-1/2 and spin-1 toroidal squared lattices, where only the
exchange interaction is working. Hence, the related Hamiltonian
can be written:
\begin{equation} \label{eq:hamiltonian}
{\cal H} = - \frac{J}{S^2} \sum_{i \sim j}^{N} \sigma_i \sigma_j,
\end{equation}
where $S$ is the spin magnitude (the spin variable $\sigma$ takes
the $2S+1$ pertinent values) and the sum only involves
nearest-neighbor pairs.

The walker performing the dynamics can move on the magnetic
lattice by hopping from one site to its nearest-neighbor and it
can also flip the relevant spin. Notice that, as in the usual
dynamics, the magnetic configuration of the lattice may remain
unaltered during one or more steps. In fact, the walker is endowed
with a non-null waiting probability and the spin-flip is only
probable. In particular, the walker is ruled by the following
(anisotropic) probability:
\begin{equation} \label{eq:probability}
{\cal P}_T(\vec{s},i,j) =
\frac{p_T(\vec{s},j)}{\sum_{\{\vec{s'}\}} \sum_{j=0}^{z_i}
p_T(j,\vec{s'})},
\end{equation}
which represents the probability that the walker, from site $i$
(with coordination number $z_i$), jumps on a n.n. site $j$, being
$\vec{s}$ the magnetic configuration attained at the step
considered. At each step, the set of all the possible magnetic
configurations (denoted as $\{\vec{s'}\}$) contains
$(2S+1)\times(z_i+1)$ elements. Hence, $\vec{s}$ represents a
particular element from \{\vec{s'}\}. Furthermore,
\begin{equation} \label{eqn:glauber}
p_T(j,\vec{s})= \frac{1}{1+e^{[\beta \Delta E_j(\textbf{s})]}}
\end{equation}
is drawn from the usual Glauber probability where $\Delta
E_j(\vec{s})$ is the energy variation consequent to the process
\cite{earlier}.

As a consequence of Eq.~\ref{eq:probability}, the random walker is
biased (BRW) towards such sites that let, by means of spin-flip, a
greater energy gain.

Notice that, as stressed in \cite{earlier}, this kind of dynamics
does violate the detailed balance condition and the equilibrium
states achieved are non-canonical. More precisely, our diffusive
dynamics can attain the actual thermodynamic behavior of the Ising
ferromagnet, even if shifted to higher temperatures. In
particular, the phase-transition occurs at temperatures
\begin{eqnarray}
T_{c}^{S=1/2} = 2.602(1)\label{eq:Tc_Mezzo}
\\
T_{c}^{S=1}=1.955(2)\label{eq:Tc}
\end{eqnarray}
larger than the canonical ones. On the other hand, the
universality class is preserved, i.e. the right critical exponents
(relatively to the 2-dimensional case) are recovered with great
accuracy. Finally, we recall that the critical temperatures
relevant to a non-diffusive dynamics are, respectively,
$T_{c}^{S=1/2} = \frac{2J}{log(1+\sqrt{2})}\approx 2.269$ and
$T_{c}^{S=1} \approx 1.695$ \cite{hoston,silva}.

\section{Geometrical Properties of spin configurations}
\label{geom}
\begin{figure}[tb]
\resizebox{0.495\columnwidth}{!}{\includegraphics{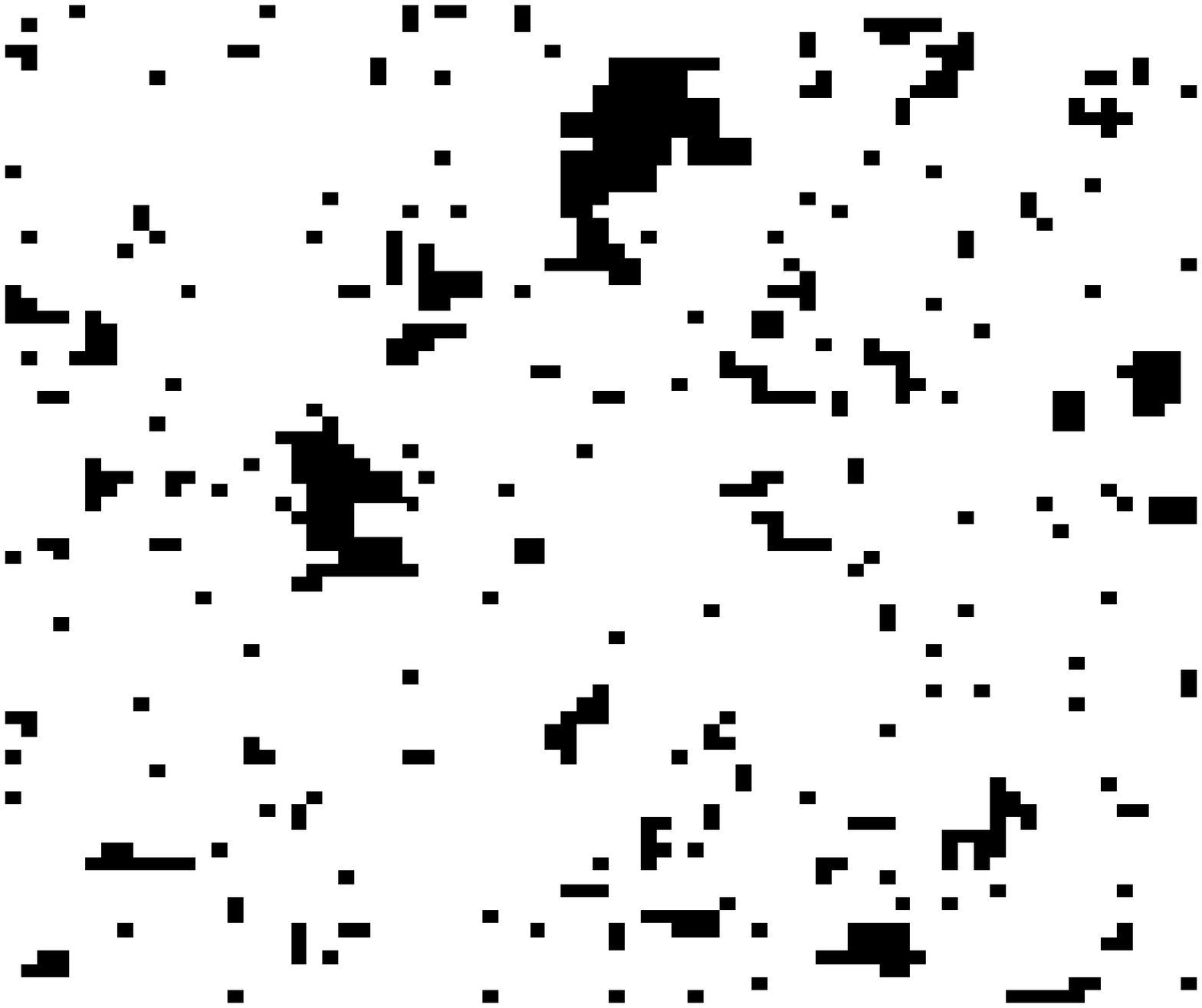}}
\resizebox{0.495\columnwidth}{!}{\includegraphics{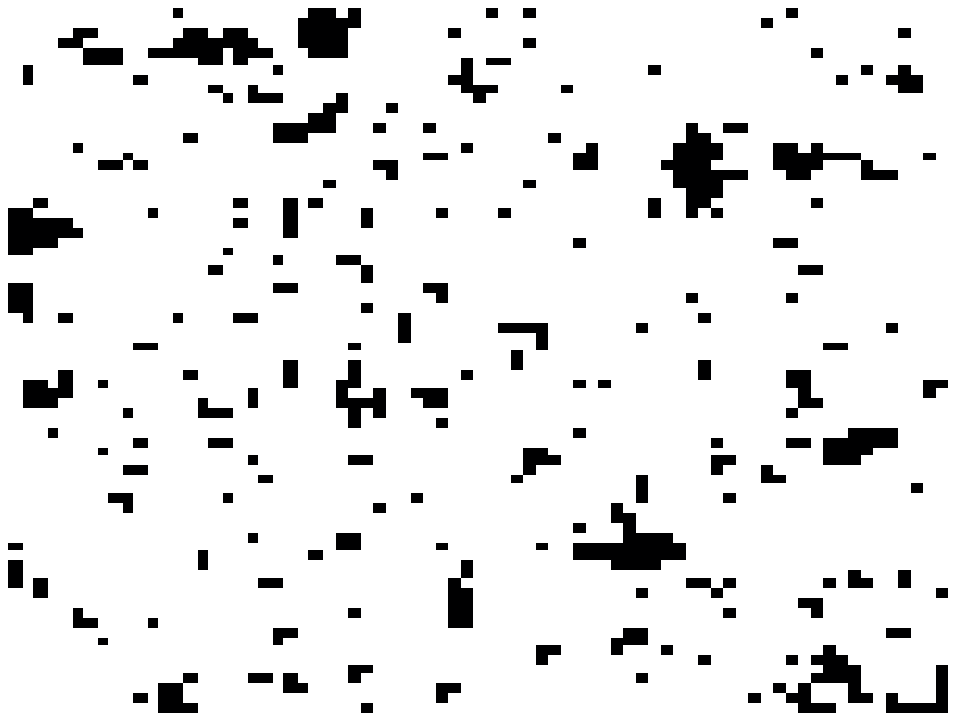}}
\caption{\label{fig:domain} These figures show a $75 \times 75$
zoom of the snapshots of typical magnetic configurations for a
$400 \times 400$ spin-1/2 Ising lattice with $\langle m \rangle =
0.84$ subject to the Glauber dynamics at $T = 2.09$ (left panel)
and to a diffusive one at $T = 2.40$ (right panel). The smallest
(black) perturbations away from the ordered (white) background
involve just a single site. Notice that the same magnetization is
attained for different temperatures and also that, due to the
diffusive character of the pertinent dynamics, the right figure
displays a greater number of black islands, though smaller.}
\end{figure}

In this section we examine the geometrical properties
featured by the magnetic pattern developing on the Ising lattice.
In particular, we want to achieve evidence that, as noticed in
\cite{earlier}, the distribution of up, down (and possibly zero)
spins on the lattice is affected by the thermal dynamics applied
to the system (Fig.~\ref{fig:domain}).

Therefore, in our analysis we ought to consider not a particular,
singular cluster, but rather the whole lattice. In this context,
we distinguish the \textit{positive cluster}, referred as the set
of sites occupied by up spins, from \textit{negative} and
\textit{null clusters}, analogously meant. The geometric analysis
of such clusters is obtained by means of the box-counting method
which, at a single blow, succeeds in taking into account the whole
magnetic arrangement.

In fact, according to the box-counting technique, an $r$-spaced
square (in general a $d$-dimensional box) grid  is superimposed
onto the object to be measured and the number of squares, $N(r)$,
overlapping a portion of the object is counted. This step is
repeated for different values of $r$. Finally, if the limit $$d_f
= \lim_{r\rightarrow 0} {\frac{logN(r)}{log(1/r)}}$$ exists, it
converges to the box-counting fractal dimension $d_f$
\cite{schroeder}. In other words, $N(r)$ scales as $r^{-d_f}$
\cite{mandelbrot}. Then, once possible sources of wrong estimates
removed \cite{ciccotti}, $d_f$ can be calculated using the linear
regression technique. More precisely, $d_f$ is assumed to be the
angular coefficient of the regression line between $logN(r)$ and
$log(1/r)$.

As remarked in \cite{ciccotti}, for an ideal fractal, such a
power-law relation holds over an infinite range of scales, while,
in real measures, some limits emerge. In particular, when dealing
with fractal images embedded in a $d$-dimensional, discrete space,
by taking the limit of box-size $r$ to zero, any curve or shape on
that space will have fractal dimension $d$. Thus, the smallest
practical box-size is $r=1$. Moreover, the range of scales can be
limited by further cutoffs such as, for example, the presence of
characteristic lengths and a limited volume of the object.

In the present case, clusters develop on a square lattice of side
$L$ and the spacing $r$ is varied amongst integer submultiples of
$L$, being $2$ its minimum value (when $r = 1$ we would simply
count the number of spins oriented in the same direction).
Actually, there also exists a constrain from above, since (below
$T_c$) the upper cutoff length for self-similarity is not the size
of the lattice, but rather the fractal correlation length $\xi$.
The latter is related to the characteristic cluster size so that,
when $r > \xi$, each covering-box contains a portion (at least one
single site) of the clusters considered. As a result, above $\xi$,
Euclidean geometry prevails and $N(r) \sim r^d$ (where $d$ is the
lattice dimension). In other words, at scales larger $\xi$, the
object to measure looks homogenous \cite{hasmy}. In the following,
the fractal correlation length relevant to negative and null
clusters will be referred as $\xi_{-1}$ and $\xi_{0}$,
respectively.
%In other words, the fractal correlation length has to do with the
%distance between fluctuations away from the ordered background:
%when $r > \xi$, each covering-box contains a portion (at least one
%single site) of the clusters considered.

In summary, our measures are carried out focusing on the proper
``fractal range'' (small scales) and, in particular,  $d_f$ is
calculated with the linear regression technique over the interval
$r \in [2, \xi]$.

It should as well be remarked that, when dealing with the very
critical region, there exist other kinds of methods
\cite{janke,coniglio,antoniou,vanderzande,vanderzande2,deng,duplantier},
specially meant, which ought to be preferred since they provide a
better reliability. On the other hand, we recall that, here, our
aim is not to measure the fractal dimension of the critical
cluster, but rather to study the behavior of the diffusive
dynamics introduced, seeking a differentiation with respect to the
non-diffusive ones, in terms of geometrical properties.

The following subsection is devoted to some general observations
concerning the evolution of the magnetic pattern, while a depth
study of the critical range is left to the second part.

\subsection{Evolution of magnetic patterns}

\begin{figure}[tb]
\resizebox{0.95\columnwidth}{!}{\includegraphics{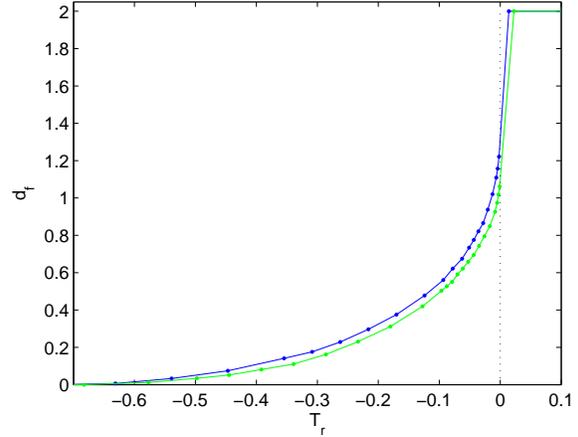}}
\caption{\label{fig:Mezzo} Fractal dimension of negative cluster
developing on an Ising lattice of linear dimension $L = 240$, made
up of spin-1/2 and subject to the diffusive (dark line) and
non-diffusive dynamics (clear line).}
\end{figure}

\begin{figure}[tb]
\resizebox{0.95\columnwidth}{!}{\includegraphics{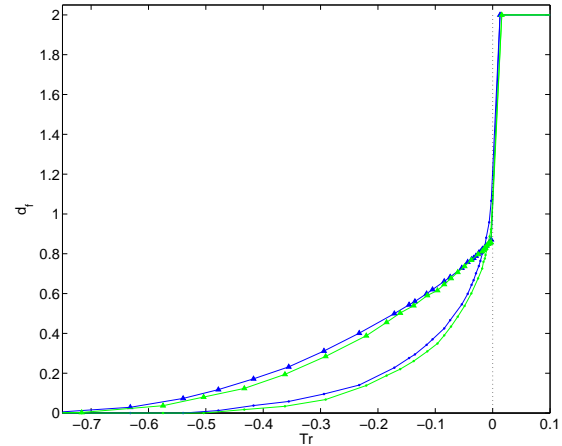}}
\caption{\label{fig:Uno} Fractal dimension of magnetic clusters
developing on an Ising lattice of linear dimension $L = 240$, made
up of spin-1 and subject to the diffusive diffusive (dark line)
and non-diffusive dynamics (clear line) Both negative ($\bullet$)
and null ($\triangle$) clusters are depicted.}
\end{figure}

In general, our measures refer to the magnetic configurations
corresponding to the steady-states of a sample initially set
ferromagnetic (${\langle m \rangle}_0 =+1$) and then heated.

Figures~\ref{fig:Mezzo} and \ref{fig:Uno} show results obtained
for the spin-1/2 and spin-1 systems, respectively; because of
different temperature scales, data pertaining to diffusive and
non-diffusive dynamics are plotted vs the reduced temperature
$(T_r=\frac{T-T_c}{T_c})$.

First of all, notice that both plots evidence a singular behavior
in $T_c$, consistent with the phase transition occurring on the
lattice. Hence, criticality also emerges from the geometry of the
spin-configuration, though observed on small scales (i.e. the
fractal range).

Also notice that, for any cluster considered, for example the one
labelled as $\nu$, we can introduce the temperature $\tau_{\nu}$
such that, if $T \leq \tau_{\nu}$ then $d_f(\nu)=0$. In other
words, for sufficiently low temperatures, negative and null spins
are still so sparse that their fractal dimension is zero.

The consistency with the magnetic evolution of the sample is not
limited to the critical point and to very low temperatures. For
example, let us consider the spin-1 system of Figure~\ref{fig:Uno}
(an analogous description also holds when $S=1/2$,
Fig.~\ref{fig:Mezzo}). As long as $\tau_{-1(0)} < T < T_c$, the
fractal dimension of negative (null) clusters increases
continuosly with the temperature and, finally, for $T > T_c$,
$d_{f}$ reaches the Euclidean dimension of the lattice. Therefore,
when the paramagnetic phase is achieved, all three kinds of
clusters share the same dimension. However, although the curves
depicted in figure both exhibit a discontinuity at $T_c$, they are
quite dissimilar. In particular, as will be discussed later, they
display a different functional dependence on $T$. Besides,
$\tau_{-1} > \tau_{0}$, due to the fact that null spins can appear
on the ferromagnetic lattice at lower temperatures, because of a
minor energy cost. For the same reason, at low temperatures, such
spins act as catalysts for the phase transition: flips in favor of
negative spins are less expensive if they engage sites adjacent
to, or occupied by null spins.

As far the non diffusive dynamics (explicitly, a Glauber dynamics
with type-writer updating), qualitatively similar plots have been
obtained. In particular, a consistent singularity is still
recorded at $T_c$ and all fractal dimensions still converge when
the reduced temperature gets positive, but, we recall, this
happens at different values of $T$ (Eq.~\ref{eq:Tc_Mezzo} and
\ref{eq:Tc}).

There are other qualitative analogies between diffusive and
non-diffusive dynamics, which concern the fractal correlation
length. Firstly, for both dynamics, $\xi$ grows slowly with the
lattice size, especially at temperatures approaching the critical
one. Consequently, in order to extend the range over which the
box-counting is applied (which means more accuracy), the size $L$
of the lattice has to be significantly enlarged. Secondly, the
self-similarity range of null clusters is very small. More
precisely, when $L=240$, $\xi_{-1}$ is about $3 \cdot 10^{-2}$
times the lattice size, while $\xi_{0}$ is about
$\frac{\xi_{-1}}{2}$. In fact, because of their negligible energy
contribution, null spins are not expected to form wide clusters
but they are arranged throughout the whole sample, amongst
positive as well as negative clusters, due to the symmetry of the
system. Therefore, while heating the sample, their density
increases and they form clusters whose typical size is so small,
that the distribution of null spins appears to be homogenous at
relatively small scales. In other words, they are more and more
likely to be found in any covering box of relatively small size.

Hitherto, we have just reported the qualitatively analogies
between the two dynamics taken into account. However, in the next
subsection, a deeper insight will highlight some interesting
differences. For the present, notice that, once the reduced
temperature fixed, the diffusive dynamics introduced generates
spin distributions which, compared with their non-diffusive
counterparts, display a greater fractal dimension
(Figs.~\ref{fig:Mezzo} and \ref{fig:Uno}). Such a difference is
related to the way each thermal dynamics deals with fluctuations
at small scales.

At very low temperatures, the magnetic patterns generated by the
two dynamics taken into account (at the same $T_r$) are
indistinguishable: perturbations away from the ordered background
are very small (mostly made up of single sites) and they are so
sparse that $d_{f}=0$. By rising the temperature, their number, as
well the correlation length, also increases and differences
emerge.

In fact, due to the bias it is endowed with, the random walker
performing the diffusive dynamics is especially concerned with
high energy sites. In particular, it spends most of its time on
border between the largest correlated regions, while fluctuations
at small scales survive quite numerous, though small-sized. On the
other hand, when a non-diffusive dynamics is applied, each site is
equivalently often considered for the updating so that
perturbations are more likely either to grow or to be destroyed.
Therefore, at small scales, correlated regions away from the
ordered background are larger and sparser (Fig.~\ref{fig:domain}).
%In other words, (non-simply connected) clusters developing on the
%lattice show, within the fractal correlation length, larger,
%though less, holes.

As a result, when covering-boxes are counted, while reducing their
size, the rate of growth of the number of boxes overlapping a
negative spin is larger in the diffusive case.

\subsection{Critical behavior}

\begin{figure}[tb]
\resizebox{0.95\columnwidth}{!}{\includegraphics{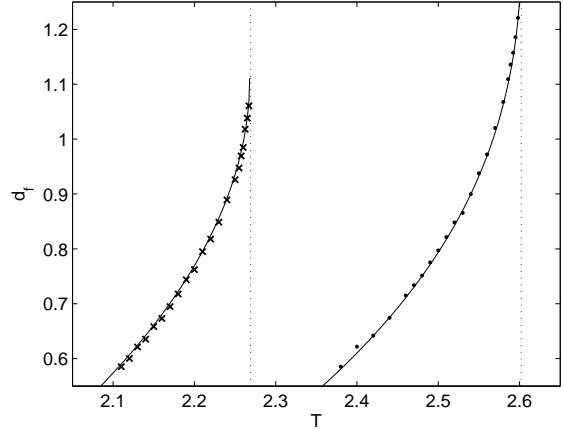}}
\caption{\label{fig:NDM}Box-counting fractal dimension for the
negative cluster relevant to the spin-1/2 Ising system on a
squared lattice. Both kind of dynamics are depicted: non-diffusive
($\times$) and diffusive ($\bullet$). The line fitting the data is
the power law of Eq.~\ref{eq:pow_law}. The vertical, dashed lines
indicate the value of the pertaining critical temperatures.}
\end{figure}

\begin{figure}[tb]
\resizebox{0.95\columnwidth}{!}{\includegraphics{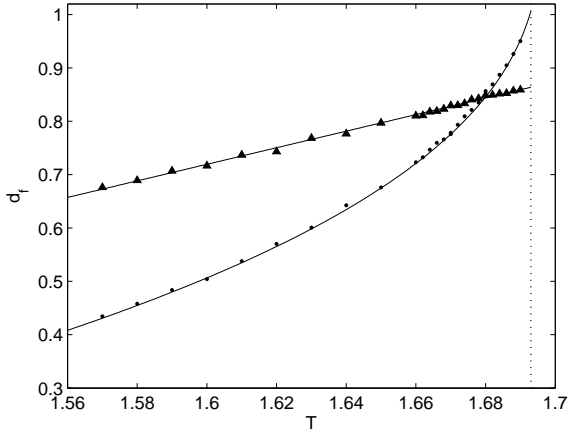}}
\caption{\label{fig:NDU}Spin-1 Ising system subject to a
non-diffusive dynamics. The box-counting dimension for the
negative ($\bullet$) and null ($\bigtriangleup$) clusters
developing on the square lattice are shown. Notice the difference
in their behavior with respect the temperature: they agree with
the power and linear laws of
Eq.~\ref{eq:pow_law},~\ref{eq:lin_law}, respectively. The
estimated critical temperature for this kind of system is
suggested by the dashed line.}
\end{figure}

\begin{figure}[tb]
\resizebox{0.95\columnwidth}{!}{\includegraphics{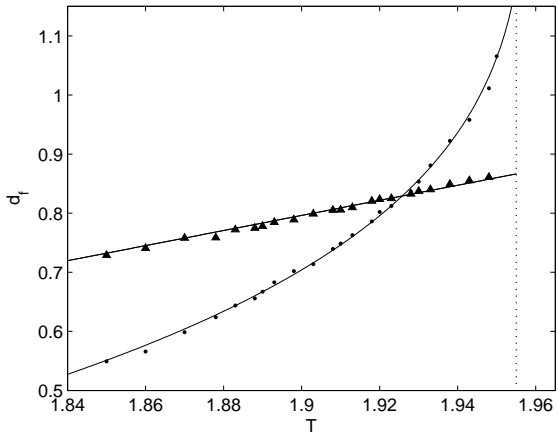}}
\caption{\label{fig:DU}Box-counting fractal dimension for the
negative ($\bullet$) and null ($\bigtriangleup$) clusters relevant
to the spin-1 Ising system on a squared lattice; the dynamics
adopted is diffusive. The lines fitting the data are a power
(Eq.~\ref{eq:pow_law}) and a linear (Eq.~\ref{eq:lin_law}) law,
respectively. The vertical, dashed line indicates the estimated
value of the critical temperature.}
\end{figure}

Now, let us focus the attention on the critical region which, in
this context, is meant as the range of temperatures approaching
(from below) $T_c$. Interestingly, for the spin-1/2 and spin-1
Ising systems taken into account (Figs.~\ref{fig:Mezzo} and
\ref{fig:Uno}), $d_{f}$ displays a functional dependence on $T$
which is affected by the kind of cluster, but by neither the
dynamics nor the spin magnitude. Specifically, the negative
cluster is described by a power law such as:
\begin{equation} \label{eq:pow_law}
d_{f} = d^{-} + A^{-}(\frac{T_c-T}{T_c})^{\kappa},
\end{equation}
while the null cluster by a linear law:
\begin{equation} \label{eq:lin_law}
d_{f} = d^{0} + A^{0}T,
\end{equation}
Due to the aforesaid range of reliability of the box-counting
technique, both Eq.~\ref{eq:pow_law} and \ref{eq:lin_law} hold
when $T < T_c$.

The value of the parameters, appearing in the previous equations,
have been obtained by fitting data plotted in
Figures~\ref{fig:NDM}-\ref{fig:DU} and are resumed in
Tab.~\ref{tab:coeff}. Notice that $T_c$ has been regarded as a
free parameter, too. Therefore, we are also able to provide a
``geometrical estimate" for the critical temperature. Actually,
such estimate well agrees, within the error, with its exact or
measured counterpart \cite{earlier}. Furthermore, it should be
underlined that $T_c$ is the most accurate among all the fitting
parameters.

\begin{table}
\begin{tabular}[htbp]{p{1.0cm}p{2.0cm}p{2.0cm}p{2.0cm}p{2.0cm}}
\cline{2-5} & \multicolumn{2}{||c||}{\bfseries{NDD}} &
\multicolumn{2}{c||}{\bfseries{DD}} \\
%\multicolumn{2}{||c||}{Non-Diffusive Dynamics} &
%\multicolumn{2}{c||}{Diffusive Dynamics} \\
\cline{2-5} & \multicolumn{1}{||c|}{$S=1/2$} &
\multicolumn{1}{c||}{$S=1$} & \multicolumn{1}{c|}{$S=1/2$} &
\multicolumn{1}{c||}{$S=1$} \\
\hline \multicolumn{1}{||c|}{\bfseries $d^{-}$} &
\multicolumn{1}{||c|}{$1.11(7)$} &
\multicolumn{1}{|c||}{$1.29(18)$} &
\multicolumn{1}{|c|}{$1.41(17)$} &
\multicolumn{1}{|c||}{$1.65(24)$} \\
\hline \multicolumn{1}{||c|}{\bfseries $A^{-}$} &
\multicolumn{1}{||c|}{$-1.98(5)$} &
\multicolumn{1}{|c||}{$-2.08(9)$} &
\multicolumn{1}{|c|}{$-2.13(8)$} &
\multicolumn{1}{|c||}{$-2.21(10)$} \\
\hline \multicolumn{1}{||c}{\bfseries $\kappa$} &
\multicolumn{1}{||c|}{$0.50(7)$} &
\multicolumn{1}{|c||}{$0.34(6)$} &
\multicolumn{1}{|c|}{$0.38(4)$}&
\multicolumn{1}{|c||}{$0.24(4)$} \\
\hline \multicolumn{1}{||c}{\bfseries $T_c$} &
\multicolumn{1}{||c|}{$2.268(4)$} &
\multicolumn{1}{|c||}{$1.698(8)$} &
\multicolumn{1}{|c|}{$2.603(6)$}&
\multicolumn{1}{|c||}{$1.958(1)$} \\
\hline \multicolumn{1}{||c}{\bfseries $d^{0}$} &
\multicolumn{2}{||c||}{$-1.76(3)$} &
\multicolumn{2}{c||}{$-1.95(3)$} \\
\hline \multicolumn{1}{||c}{\bfseries $A^{0}$} &
\multicolumn{2}{||c||}{$1.55(2)$} &
\multicolumn{2}{c||}{$1.33(2)$} \\
\hline
\end{tabular}

\caption{Fit coefficients referring to Eq.~\ref{eq:pow_law} and
\ref{eq:lin_law}, pertaining to the non-diffusive dynamics
(\textbf{NDD}) and the diffusive dynamics introduced
(\textbf{DD}).} \label{tab:coeff}
\end{table}

As far the other parameters, while $d^{-}$ provides a rough
estimate for the fractal dimension of the critical negative
cluster, we are especially interested in $\kappa$ and $A^0$, since
they do characterize the evolution of spin arrangements. Indeed,
as shown by Tab.~\ref{tab:coeff}, $\kappa$ and $A^0$ depends on
both the dynamics adopted and the spin magnitude $S$. This means
that, once the dynamics (diffusive or non-diffusive) selected, the
negative clusters pertaining to spin-1/2 and spin-1 systems,
respectively, show a different geometric evolution. Not only, the
geometric evolution of a particular cluster is also
dynamic-sensitive.

This is really a non-trivial result since, from a thermodynamic
point of view, the dynamics introduced actually recovers the
canonical critical behavior. More precisely, the critical exponent
$\alpha$, $\beta$, $\gamma$ have revealed to be the same, in the
spin-1/2, as well as in the spin-1, 2-dimensional systems
\cite{earlier}. Therefore, as far the critical behavior is
concerned, the very effects of our dynamics have to be tracked
down in the geometry of magnetic patterns, rather than in the
thermodynamic quantities.

In particular, the exponent $\kappa$ pertinent to the diffusive
dynamics is smaller, which means that, as $T_r \rightarrow 0^-$,
the rate of growth of the fractal dimension of the related
negative cluster is larger. Such phenomena are consistent with the
above described effects of the diffusive, biased character of the
dynamics introduced. Moreover, as you can see from
Tab.~\ref{tab:coeff}, the same holds not only when $S=1/2$, but
also when $S=1$. In fact, the fitting parameters of
Eq.~\ref{eq:pow_law} display the same inequalities when diffusive
and non-diffusive dynamics are compared.

On the other hand, once the thermal dynamics selected, the spin-1
system shows, with respect the $S=1/2$ case, smaller exponents
$\kappa$ and larger parameters $d^-$. The reason is that, as
previously underlined, when $T < T_c$, the presence of null spins
makes the negative ones more likely to occur. In fact, due to
their energetic neutrality, null spins can be spread throughout
the whole lattice, at a relatively small energy cost. Now, if
$\sigma_i=0$, transitions towards $\sigma_i=-1$ or $\sigma_j=-1$
(where $i \sim j$) show significantly reduced energy costs and are
therefore more likely to happen. As a result, not only negative
spins are more uniformly distributed, but their number rapidly
increases as $T \rightarrow T_c$.

As far the null cluster, the linear law evinced reveals that,
interestingly, its evolution is indeed qualitatively different
with respect the negative cluster. More precisely, its growing
dynamics is slower (especially when the diffusive dynamics is
applied). However, comparisons in this sense should be quite
careful since, we recall, the related fractal lengths are not the
same.

Some remarks are in order now. The values of the fitting
parameters, especially the exponent $\kappa$, are quite sensitive
to the measures of the fractal dimension $d_{f}$, just nearby the
critical point. This explains the quite large errors affecting the
exponents $\kappa$, which may also not forbid their overlapping.
On the other hand, the power and linear laws evinced, as well as
the differences displayed by the diffusive dynamics, still hold
within the error.

\begin{figure}[tb]
\resizebox{0.95\columnwidth}{!}{\includegraphics{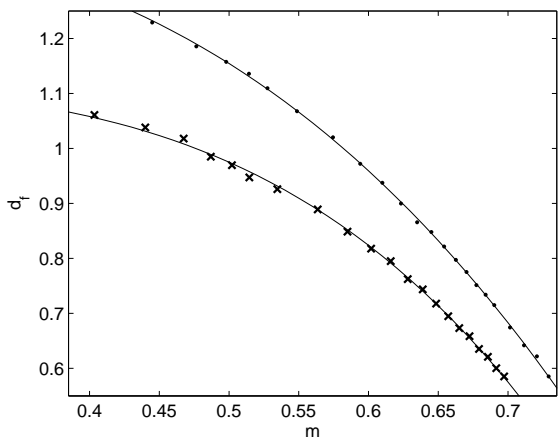}}
\caption{\label{fig:df_m}Box-counting fractal dimension for the
negative cluster obtained with a diffusive ($\bullet$) and
non-diffusive ($\times$) dynamics applied to the spin-1/2 Ising
system. The box-counting dimension is depicted as a function of
the total magnetization $m$ and the lines fitting the data can be
derived from (Eq.~\ref{eq:pow_law}). In particular, $d_f$ scales
as $m^{\frac{\kappa}{\beta}}$.}
\end{figure}

Finally, notice that, in the analysis carried on, we could have
considered $d_f$ as a function of the total magnetization $m$
rather than as a function of the reduced temperature $T_r$.
However, this change would not lead to any significant
differences. In fact, the power and linear laws of
Eq.~\ref{eq:pow_law} and \ref{eq:lin_law} would still hold, though
their coefficients would be properly modified. This is a
consequence of the aforesaid analogy between the thermodynamic
critical behavior displayed by diffusive and non-diffusive
dynamics. In particular, the two different dynamics would still
provide different exponents $\kappa$, though rescaled by a factor
$\beta$ with respect to the original ones (Fig.~\ref{fig:df_m}).

\section{Conclusions}
\label{concl}The thermal diffusive dynamics introduced and
analyzed in \cite{earlier} has been further inspected, focusing
the attention on the geometrical properties featured by clusters
developing on the magnetic lattice. In particular, by means of the
box-counting method, we measured and compared the fractal
dimension of magnetic patterns generated by our diffusive dynamics
and by a non-diffusive one. We have also taken into account either
spin-1/2 and spin-1 systems.

As expected, in any case, when $T > T_c$ each cluster (positive,
negative and, possibly, null) shows the dimension of the lattice
itself. In fact, when the paramagnetic phase is reached, each kind
of cluster is likely to overlap any covering-box, due to the
discreteness of the lattice and to a limited number of possible
spin-states. On the other hand, when $T \leq T_c$ a much more
interesting scenario appears.

In general, the fractal dimension $d_{f}$ of clusters displays a
dependence on $T$ consistent with the thermodynamic evolution of
the magnetic lattice. In particular, in $T_c$, a singularity
evidences the occurrence of the phase transition. Hence,
criticality also emerges from the small-scales fractal geometry of
magnetic patterns. Besides, as $T_r \rightarrow 0^-$, the fractal
dimension of the negative cluster depends on $T_r$ through a power
law. The related fitting coefficients provide not only good
estimates of the critical temperature, but they also highlight
some interesting differences between the diffusive and
non-diffusive dynamics. In particular, the former, as a
consequence of its diffusive, biased character, generates a less
sparse distribution of negative spins and then, the related
fractal dimension is larger. Our data also suggest the exponent
characterizing the diffusive dynamics to be smaller, which means a
larger rate of growth for $d_f$, as approaching $T_c$.

Therefore, the very effects of the diffusive dynamics introduced
lie in the geometrical, rather than thermodynamic \cite{earlier},
critical behavior of the system.

Such effects occur for both the spin-1/2 and spin-1 systems
considered. Interestingly, in the latter case, negative clusters
display even smaller exponents, while null clusters show a
distinct behavior. In particular, their fractal dimension depends
linearly on $T$. However, it should be underlined that the fractal
range pertinent to negative and null clusters is different, being
smaller for the latter. Such differences has to be attributed to
the special role played by the state $\sigma = 0$. In fact, as
energetically neutral, it is prevented by forming wide domains,
while, at sufficiently large temperatures, it is likely to be
found throughout the whole lattice. We argue that it would be
worth studying to what extent such a gap among spin states affects
the BRW behavior as well as the shape of clusters. For example,
one could take into account systems like the $q$-state Potts model
or the spin-S ($S > 1$) Ising model, where a
different symmetry among spin-states is present.% \cite{forth}.

\end{document}